\pdfoutput=1
\documentclass[11pt]{article}

\usepackage[final]{acl}
\usepackage{multirow}
\usepackage{times}
\usepackage{latexsym}

\usepackage[T1]{fontenc}
\usepackage[utf8]{inputenc}

\usepackage{microtype}

\usepackage{inconsolata}

\usepackage{graphicx}
\usepackage{xspace}
\usepackage{amsmath}
\usepackage{enumitem}
\usepackage{booktabs}

\usepackage{lineno}
\nolinenumbers 

\newcommand{\Ours}[0]{Molar\xspace}

\title{Molar: Multimodal LLMs with Collaborative Filtering Alignment for Enhanced Sequential Recommendation}

\author{%
  Yucong Luo, Qitao Qin, Hao Zhang, Mingyue Cheng, Ruiran Yan, Kefan Wang, Jie Ouyang 
   \\
   University of Science and Technology of China \\
   State Key Laboratory of Cognitive Intelligence \\ 
   Hefei, Anhui, China \\
  \texttt{\{prime666,qqt,zh2001,yanruiran,wangkefan,ouyang\_jie\}@mail.ustc.edu.cn} \\ \texttt{mycheng@ustc.edu.cn}\\
  }

\begin{document}
\maketitle

\begin{abstract}
Sequential recommendation (SR) systems have evolved significantly over the past decade, transitioning from traditional collaborative filtering to deep learning approaches and, more recently, to large language models (LLMs). While the adoption of LLMs has driven substantial advancements, these models inherently lack collaborative filtering information, relying primarily on textual content data neglecting other modalities and thus failing to achieve optimal recommendation performance. To address this limitation, we propose \textbf{\Ours}, a \textbf{M}ultim\textbf{o}dal large \textbf{la}nguage sequential \textbf{r}ecommendation framework that integrates multiple content modalities with ID information to capture collaborative signals effectively. 
\textbf{\Ours} employs an MLLM to generate unified item representations from both textual and non-textual data, facilitating comprehensive multimodal modeling and enriching item embeddings. Additionally, it incorporates collaborative filtering signals through a post-alignment mechanism, which aligns user representations from content-based and ID-based models, ensuring precise personalization and robust performance. By seamlessly combining multimodal content with collaborative filtering insights, \textbf{\Ours} captures both user interests and contextual semantics, leading to superior recommendation accuracy. Extensive experiments validate that \textbf{\Ours} significantly outperforms traditional and LLM-based baselines, highlighting its strength in utilizing multimodal data and collaborative signals for sequential recommendation tasks.
The source code is available \footnote{https://anonymous.4open.science/r/Molar-8B06/}.
\end{abstract}

\section{Introduction}

In the era of information overload, recommender systems have become essential tools to filter information through various online applications, including e-commerce, advertising, and online video platforms \cite{resnick1997recommender, koren2008factorization}. Among these systems, sequential recommendation (SR) methods \cite{wang2019sequential, zhou2018deep, kang2018self,cheng2021clue} have gained prominence due to their ability to capture dynamic user interests more effectively than traditional collaborative filtering techniques. Mainstream SR approaches are mainly based on ID-based deep learning strategies \cite{koren2009matrix, goldberg1992using}, such as matrix factorization and deep sequence neural networks~\cite{cheng2022towards}. These methods encode users and items as unique identifiers, using historical interaction data to learn sequential behavioral patterns. 
\begin{figure}[t]
    \centering
    \includegraphics[width=0.5\textwidth]{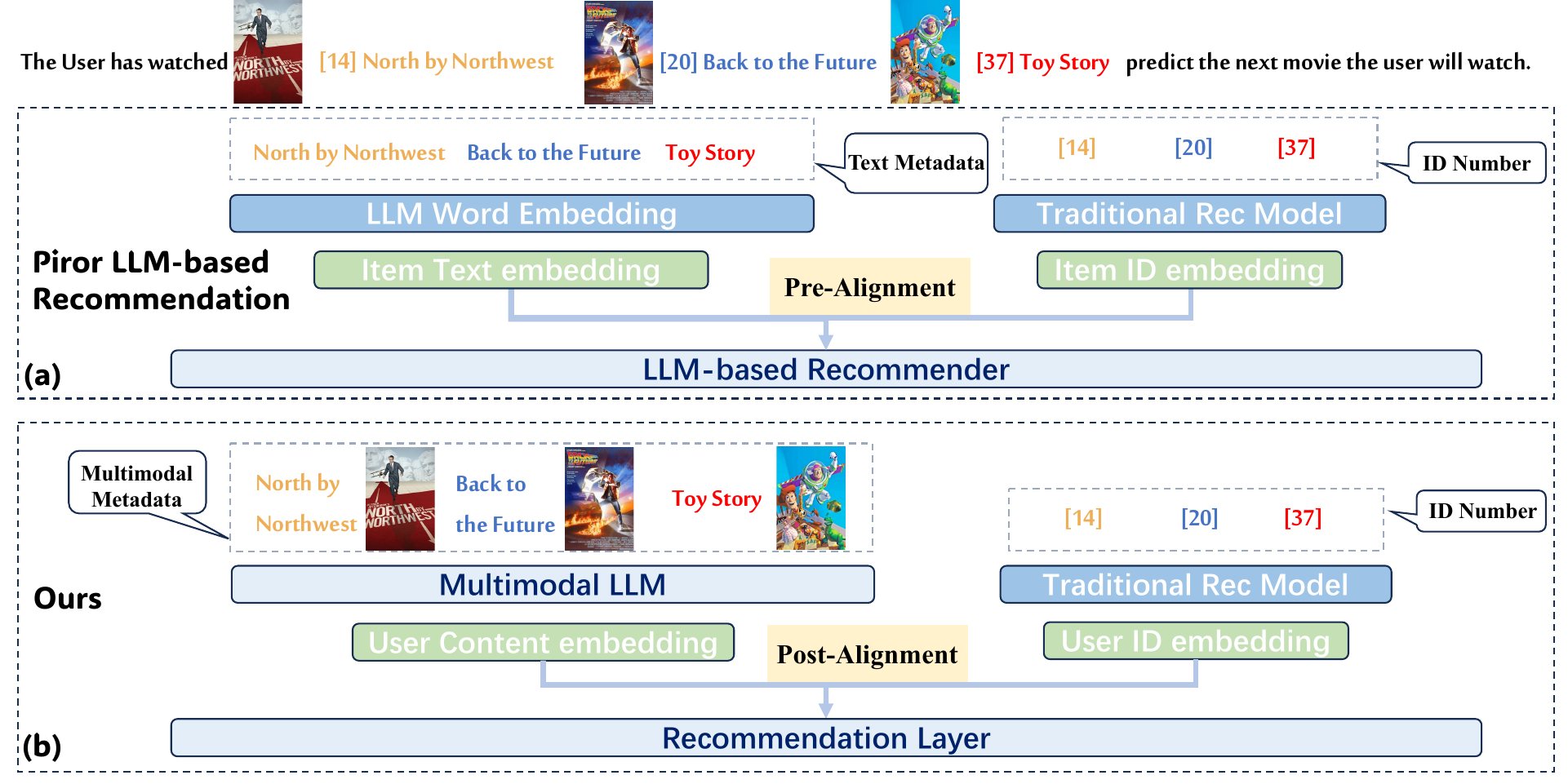} % 调整宽度
\caption{\textbf{Comparison of LLM-based recommendation methods and our \Ours}. (a) Existing methods prematurely integrate ID and text modalities into the LLM, leading to limited utilization of multimodal content features. (b) Our approach first processes text and non-text modalities through the LLM to generate rich multimodal representations and then incorporates ID information via post-alignment, ensuring a better balance between multimodal content and collaborative signals.}
    \label{fig:0_motivation}
\end{figure}

Recent advances in large language models (LLMs) \cite{zhao2023survey,luo2024unlocking} have opened new possibilities for sequential recommendation. With their powerful sequence modeling and multimodal understanding capabilities, LLMs have been explored in two main directions. The first approach \cite{zhang2023collm, ren2024representation, friedman2023leveraging} reframes SR as a natural language processing task, allowing LLMs to interpret user sequences and generate recommendations based on their language understanding. The second approach \cite{ning2024user, liao2023llara} combines LLMs with traditional SR models by integrating ID and text modalities into the LLM backbone at an early stage, as shown in Figure \ref{fig:0_motivation}a. However, these methods face critical limitations: (1) inadequate integration of multimodal features, leading to information loss from non-textual modalities or misalignment between textual and visual features; and (2) suboptimal utilization of traditional SR models, where early fusion of ID information can cause LLMs to learn shortcuts, overshadowing collaborative filtering signals. These challenges result in a failure to fully exploit the multimodal features of items and the potential of collaborative filtering in SR.

To address these issues, this paper introduces the \textbf{\Ours}, a \textbf{M}ultim\textbf{o}dal large \textbf{la}nguage sequential \textbf{r}ecommendation framework. We make three main contributions in this paper.
\textit{First}, we propose a Multimodal Item Representation Model (MIRM) based on a multimodal large language model (MLLM) to extract item features from both textual and non-textual modalities. By fine-tuning MIRM on multimodal data, we ensure robust and consistent item embeddings.  
\textit{Second}, we design a Dynamic User Embedding Generator (DUEG) that leverages these item embeddings to model user interests and predict future behaviors. This allows for effective user modeling in complex multimodal scenarios.  
\textit{Third}, we introduce a post-alignment contrastive learning mechanism that aligns collaborative signals from ID-based and content-based models, preserving the strengths of both representations. This mechanism ensures semantic alignment between user embeddings while enhancing recommendation accuracy.

Experiments conducted on multiple datasets demonstrate the effectiveness of \Ours. The framework significantly outperforms traditional SR models and state-of-the-art LLM-based methods, achieving superior results in recommendation accuracy and robustness. Our results show that \Ours captures user interests more comprehensively by combining multimodal content and collaborative filtering signals, leading to consistent performance improvements across diverse scenarios.

Our contributions are summarized as follows:
\begin{itemize}[left=0.em, itemsep=-5pt, topsep=5pt]
\item We propose a Multimodal Item Representation Model (MIRM) to extract robust item embeddings by leveraging multimodal content, including both textual and non-textual modalities, ensuring comprehensive item feature modeling.

\item We design a Dynamic User Embedding Generator (DUEG) to effectively model user preferences using multimodal item embeddings, enabling dynamic and accurate user interest prediction.

\item We introduce a post-alignment contrastive learning mechanism to integrate collaborative filtering signals from ID-based and content-based models, preserving their complementary strengths and enhancing recommendation performance.

\end{itemize}

\section{Related Work}
\paragraph{LLM-Based Recommendation.}
The success of LLMs such as GPT4 \cite{openai2024gpt4technicalreport} and LLaMA \cite{grattafiori2024llama3herdmodels} has sparked extensive exploration of their application in recommendation systems. Firstly, LLMs are used to enhance user or item information, such as aligning semantic spaces for user and item profiling or generating training signals for cold-start items \cite{xi2024towards, ren2024representation, zhang2024spar}. Secondly, some studies convert recommendation data into a conversational format, leveraging LLMs' instruction-following capabilities to predict user behavior \cite{friedman2023leveraging, bao2023tallrec, zhang2023recommendation}. Lastly, LLMs are adapted for recommendation tasks by combining ID-based item embeddings with textual features for hybrid prompting or using them for multi-class classification and regression for rating prediction \cite{ning2024user, liao2023llara}. Although these methods demonstrate the potential of LLMs, improvements over traditional recommendation models remain limited. Prior methods either overlook traditional ID-based models by focusing only on text or introduce ID modalities too early, reducing the effectiveness of collaborative filtering signals during LLM training. Unlike these approaches, we propose a post-alignment mechanism to fuse ID-based collaborative information later in the process, preserving LLMs' world knowledge while retaining essential collaborative information.

\begin{figure*}[h] 
    \centering
    \includegraphics[width=1\textwidth]{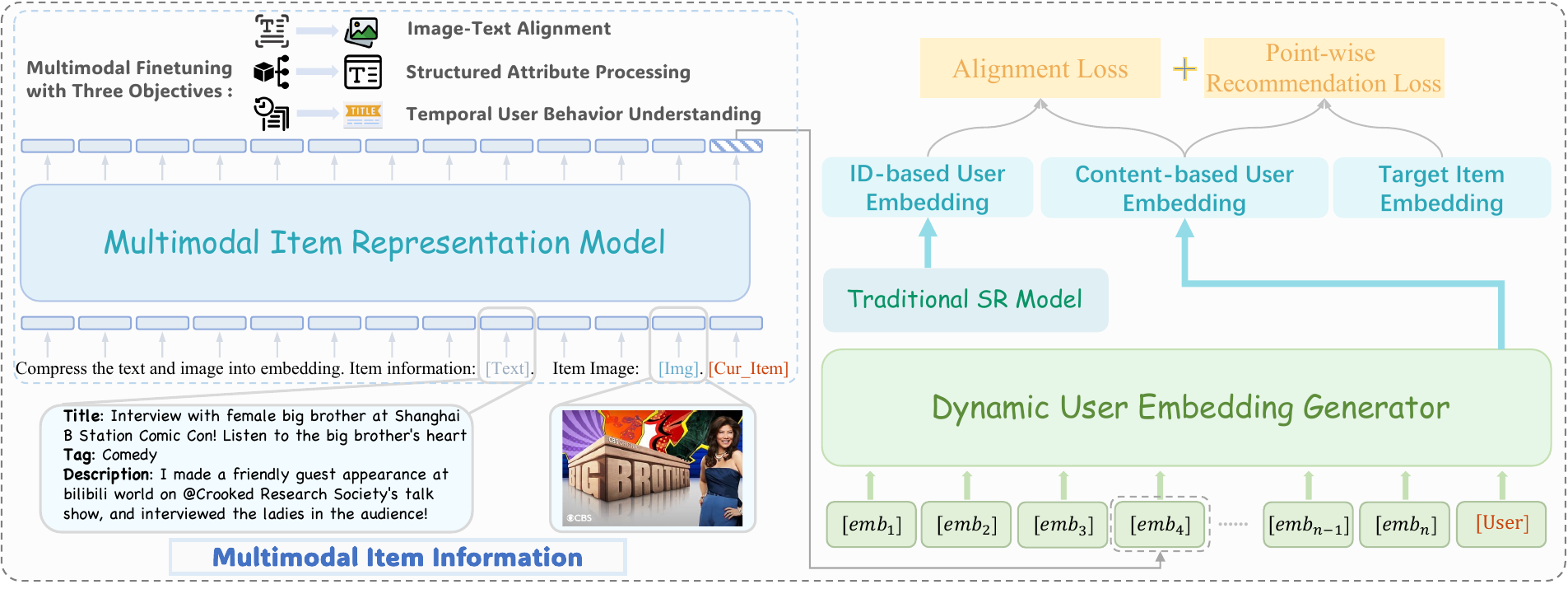} 
    \vspace{-0.3in}
    \caption{
\textbf{Illustration of the \Ours framework.} The Multimodal Item Representation Model (MIRM) processes multimodal item information to generate item embeddings, while the Dynamic User Embedding Generator (DUEG) models user embeddings based on interaction histories for next-item prediction. First, MIRM is fine-tuned for multimodal feature alignment. Then, a joint optimization framework integrates ID-based and content-based user embeddings using a contrastive learning mechanism to enhance recommendation performance. } 
    \label{fig:1_framework} 
    \vspace{-0.2in}
\end{figure*}
\paragraph{Multimodal Large Language Models.}
Recent advancements \cite{peng2023kosmos, zhang2024mm, yin2024survey} in multimodal pre-training have significantly improved task performance but at the cost of increased computational demands. To address this, researchers are leveraging pre-trained unimodal models, particularly large language models (LLMs) \cite{kasneci2023chatgpt}, to develop Multimodal Large Language Models (MLLMs) that integrate language with other modalities. The primary challenge lies in achieving effective model collaboration, with a focus on aligning modalities and understanding human intent. MLLMs like GPT-4 \cite{openai2024gpt4technicalreport} and Gemini \cite{team2023gemini} have demonstrated exceptional capabilities in multimodal comprehension. Some studies \cite{wang2024qwen2, lu2024deepseek, zhang2024notellm}  have concentrated on integrating LLMs with visual encoders. Inspired by these prior studies, we have developed \Ours, which utilized MLLMs to align and fuse multimodal information to enhance sequential recommendation.

\section{Methods}

\subsection{Problem Formulation}
We tackle the task of sequential recommendation, where the goal is to predict the next item \(I_{n+1}\) that a user \(u \in U\) is likely to interact with, given their historical interaction sequence \(H_u = \{I_1, I_2, \dots, I_n\}\) arranged in chronological order. Each item \(I_i \in I\) comes with multimodal information, such as titles, textual descriptions, and images. Our approach leverages this multimodal information to improve the prediction accuracy of the next interaction.

\subsection{\Ours Overview}
Traditional recommendation systems based on Large Language Models (LLMs) often suffer from inefficiencies when handling extensive user histories, as transforming these histories into lengthy token sequences results in high computational costs and slower inference speeds. To address these challenges, we propose \textbf{\Ours}, a decoupled framework that separates the modeling of items and users for more efficient processing. This separation allows tailored-modeling strategies for each component, improving computational efficiency and recommendation accuracy.
Our framework is composed of two key modules: the {Multimodal Item Representation Model (MIRM)} and the {Dynamic User Embedding Generator (DUEG)}. MIRM is designed to compress multimodal features into compact embeddings, mitigating the computational burden of lengthy token sequences. DUEG then utilizes these embeddings to build user representations that capture dynamic user preferences. Together, these modules enable effective multimodal feature modeling and user preference prediction, setting the foundation for robust sequential recommendation.

\subsection{Multimodal Item Representation Model.}

To effectively extract and encode item features, we introduce the {Multimodal Item Representation Model (MIRM)}, denoted as \(f_I\). This encoder leverages a multimodal large language model (MLLM) to combine textual descriptions and images into a unified embedding representation. Although MLLMs excel in understanding text and images, their direct application to feature extraction remains limited. To address this, we append a special identifier, \texttt{[Cur\_Item]}, to the end of each item's description, guiding the model to focus on extracting relevant features.

The process begins by merging an item's textual and image attributes into a unified description \(L\), augmented with a prompt to enhance model comprehension. This description is tokenized and processed by the MLLM, with \texttt{[Cur\_Item]} appended at the end of the token sequence \(\{t_1, t_2, \dots, t_m, \texttt{[Cur\_Item]}\}\). The model's hidden state corresponding to \texttt{[Cur\_Item]} is extracted as the item's embedding:
\begin{equation}
\langle emb_i \rangle = f_I(txt_i, img_i),
\end{equation}
where \(\langle emb_i \rangle\) is the embedding of item \(I_i\), \(txt_i\) is the textual description, \(img_i\) is associated image.

To enhance the quality of representations, MIRM undergoes multimodal fine-tuning with three complementary objectives:

\paragraph{Image-Text Alignment.}
Aligns visual features with textual descriptions, using item images to generate detailed descriptions.
This alignment ensures that the model captures meaningful relationships between visual content and textual context, improving its ability to interpret multimodal data cohesively.  The fine-tuning data for this objective (\textbf{Image-Text, IT}) uses item images as input and produces detailed textual descriptions as output.

\paragraph{Structured Attribute Processing.}
Converts structured metadata (e.g., price, material) into natural language descriptions for comprehensive feature encoding.
This process allows the model to integrate diverse item attributes into a unified representation, enhancing its flexibility to handle heterogeneous data types.
The fine-tuning data (\textbf{Structured Attributes, SA}) uses item titles and metadata (e.g., price, material, size, color) as input to generate detailed textual descriptions as output.

\paragraph{Temporal User Behavior Understanding.}
Captures temporal dynamics by predicting future items from historical interactions using multimodal inputs.
This objective helps the model learn sequential patterns in user behavior, enabling it to better adapt to dynamic user preferences over time. 
The fine-tuning data (\textbf{User Behavior, UB}) consists of historical item interactions (descriptions and images) as input, with predicted next items as output. 

These objectives collectively enable MIRM to produce robust, multimodal embeddings that integrate seamlessly into the subsequent user modeling process, bridging item representation with user preference prediction.

\subsection{Dynamic User Embedding Generator}
Building on the item embeddings generated by MIRM, we design the {Dynamic User Embedding Generator (DUEG)}, denoted as \(f_U\). This module constructs dynamic user representations based on their historical interactions, effectively capturing evolving preferences. Unlike MIRM, DUEG simplifies the structure by removing the word embedding layer from the MLLM, retaining the pretrained parameters for efficient embedding processing.

Given a user’s historical interaction sequence \(H_u = \{I_1, I_2, \dots, I_n\}\), MIRM transforms each item \(I_i\) into an embedding \(emb_i\). These embeddings are then processed by DUEG, which incorporates a special \texttt{[User]} token to represent the user’s dynamic preferences. This approach enables DUEG to predict the next likely item \(I_{n+1}\) based on past interactions, formalized as:
\begin{equation}
E_{u} = f_U(emb_1, emb_2, \dots, emb_n),
\end{equation}
where \(E_u\) is the user embedding.

To optimize both MIRM and DUEG, we introduce two loss functions:

\paragraph{Point-wise Recommendation Loss.}  
A binary cross-entropy loss distinguishes between positive and negative samples, encouraging accurate next-item predictions:
\begin{equation}
\mathcal{L}_{\text{bce}} = - \left( y \cdot \log(x) + (1 - y) \cdot \log(1 - x) \right),
\end{equation}
where \(y\) is the label vector, and \(x\) represents predicted logits for positive and negative samples. Details of this loss can be found in the Appendix \ref{rec_loss}.

% 这个 loss 比较重要，多写点
\paragraph{Alignment Loss.}  
% A contrastive learning objective aligns user embeddings derived from collaborative filtering (ID-based) and content-based models:

To bridge content-based and ID-based representations, we introduce a post-alignment mechanism that uses a contrastive learning objective for both embeddings after DUEG processes multimodal content. This prevents premature integration of collaborative filtering into the LLM, ensuring essential collaborative information is preserved. The contrastive learning objective is defined as follows:

\begin{small}
   \begin{align}
% \hspace{-0.11in}
\mathcal{L}_{align} = - \frac{1}{|U|}  \sum_{u=1}^{|U|} & \bigg(  \log \frac{\exp(s({E}_u^{id}, {E}_u^{con})/\tau)}{\sum\limits_{j \in K} \exp(s({E}_u^{id}, {E}_j^{con})/\tau)} \notag \\
& + \log \frac{\exp(s({E}_u^{con}, {E}_u^{id})/\tau)}{\sum\limits_{j \in K} \exp(s({E}_u^{con}, {E}_j^{id})/\tau)} \bigg),
\end{align} 
\end{small}
where \(\tau\) is the temperature parameter, \(K\) is the set of comparative instances containing one positive and \(K\)-1 negative examples. \(s(\cdot,\cdot)\) denotes the cosine similarity function, and \({E}^{id}\) and \({E}^{con}\) are the ID-based user embeddings from a traditional sequential recommendation model and the content-based embeddings from DUEG, respectively.

The final training objective combines two losses:
\begin{equation}
\mathcal{L}_{\text{total}} = \mathcal{L}_{\text{bce}} + \alpha \cdot \mathcal{L}_{\text{align}},
\end{equation}
where \(\alpha\) balances their contributions.

By integrating multimodal item features with collaborative signals, DUEG enhances accuracy and robustness in user preference modeling, enabling a seamless transition to recommendation generation.

\section{Experiments}
In this section, we evaluate our proposed framework, \Ours, on three real-world datasets and compare it against several baselines, including traditional sequential recommender models and state-of-the-art LLM-based methods. To assess the effectiveness of \Ours, we conduct a comprehensive analysis addressing four research questions. Additionally, we investigate the impact of different DUEGs and the various input data modalities on the results.Furthermore, we perform ablation studies to investigate the impact of fine-tuning strategies and post-alignment, as well as evaluate the performance differences between our LLM-based user modeling approach and alternative user representation methods. The following research questions are explored:

\begin{itemize}[left=0.em, itemsep=-5pt, topsep=5pt]
    \item \textbf{RQ1}: How does \Ours perform compared with traditional sequential recommender models and LLM-based methods?  
    \item \textbf{RQ2}: What are the differences in performance between the LLM DUEG and alternative DUEGs for user representation?  
    \item \textbf{RQ3}: How do different data modalities impact the performance of \Ours?
    \item \textbf{RQ4}: How does the post-alignment model affect the performance of \Ours?
    \item \textbf{RQ5}: How does fine-tuning the MIRM combined with post-alignment training influence the overall performance of \Ours?  
\end{itemize}

\subsection{Experimental Settings}

\subsubsection{Datasets and Evaluation Metrics}

\begin{itemize}[left=0.em, itemsep=-5pt, topsep=5pt]
    \item \textbf{Amazon} \footnote{\url{https://jmcauley.ucsd.edu/data/amazon/}}: Collected from the Amazon cloth online shopping platform.
    \item \textbf{PixelRec} \footnote{\url{https://github.com/westlake-repl/PixelRec}}: An image dataset for recommender systems with raw pixels and text.
    \item \textbf{MovieLens} \footnote{\url{https://grouplens.org/datasets/movielens/}}: A commonly-used movie recommendation dataset that contains user ratings.
\end{itemize}

\begin{table}[t]
    \centering
    \caption{Statistics of Datasets.}
    \label{tab:1_data_statistics}
    {\footnotesize
    \begin{tabular}{lrrr}
    \toprule
    Dataset   & Amazon  & PixelRec & MovieLens  \\
    \midrule
    \# User    & 993,087      & 50,000 & 6,040                   \\
    \# Item     & 301,312         & 82,864 & 3,706                    \\
    \# Interaction  & 8,813,442      & 989,476 & 1,000,209                \\
    \bottomrule
    \end{tabular}
    }
\vspace{-0.15in}
\end{table}

\begin{table*}[h]
\centering
\caption{\textbf{Performance comparison of \Ours with baseline models.} The underlined values indicate the best and second-best results across all models. The abbreviations N and R represent Normalized Discounted Cumulative Gain (NDCG) and Recall, respectively. Statistically significant improvements are marked with * ($p$-value $<< 0.05$). Overall, \Ours consistently achieves superior performance across all datasets, demonstrating its effectiveness in leveraging multimodal and collaborative filtering features.}
\vspace{-0.1in}
\label{tab:2_main_results}
\scalebox{0.68}{
\begin{tabular}{@{}llcccccccccccc@{}}
\toprule
\multirow{2}{*}{} & \multirow{2}{*}{Methods} & \multicolumn{4}{c}{Amazon*} & \multicolumn{4}{c}{PixelRec*} & \multicolumn{4}{c}{Movielens*} \\
\cmidrule(lr){3-6} \cmidrule(lr){7-10} \cmidrule(lr){11-14}
 &  & N@10 & N@20 & R@10 & R@20 & N@10 & N@20 & R@10 & R@20 & N@10 & N@20 & R@10 & R@20 \\
\midrule
\multirow{4}{*}{Traditional} 
 & FPMC    & 0.1037 & 0.1059 & 0.1152 & 0.1232 & 0.0107 & 0.0129 & 0.0191 & 0.0290 & 0.0907 & 0.1129 & 0.1708 & 0.2756 \\
 & GRU4Rec & 0.1029 & 0.1054 & 0.1107 & 0.1190 & 0.0109 & 0.0127 & 0.0189 & 0.0284 & 0.0828 & 0.1081 & 0.1657 & 0.2664 \\
 & SASRec  & 0.1080 & 0.1105 & 0.1188 & 0.1281  & 0.0131 & 0.0149 & 0.0207 & 0.0311 & 0.1116 & 0.1395 & 0.2137 & 0.3245 \\
 & DuoRec & 0.1281 & 0.1342 & 0.1406 & 0.1616 & 0.0147 & 0.0181 & 0.0241 & 0.0362 & 0.1530 & 0.1790 & 0.2704 & 0.3738 \\
\midrule
\multirow{3}{*}{Content-based} 
 & SASRec$_{Bert}$ & 0.1116 & 0.1130 & 0.1275 & 0.1365 & 0.0131 & 0.0161 & 0.0238 & 0.0357 & 0.1172 & 0.1465 & 0.2244 & 0.3407 \\
 & SASRec$_{Vit}$ & 0.1142 & 0.1187 & 0.1237 & 0.1373 & 0.0126 & 0.0155 & 0.0211 & 0.0317 & 0.1204 & 0.1499 & 0.2295 & 0.3481  \\
 & SASRec$_{Bert+Vit}$ & 0.1164 & 0.1179 & 0.1308 & 0.1437 & 0.0136 & 0.0167 & 0.0210 & 0.0315 & 0.1258 & 0.1567 & 0.2382 & 0.3599 \\
\midrule
\multirow{2}{*}{LLM-based} 
 & CoLLM & \underline{0.1298} & 0.1344 & 0.1388 & 0.1602 & 0.0173 & 0.0213 & 0.0296 & 0.0444 & \underline{0.1658} & 0.1880 & 0.2895 & \underline{0.4058}  \\
 & HLLM & 0.1285 & \underline{0.1351} & \underline{0.1457} & \underline{0.1668} & \underline{0.0189} & \underline{0.0232} & \underline{0.0352} & \underline{0.0528} & 0.1652 & \underline{0.1933} & \underline{0.2920} & 0.4037  \\
\midrule
\multirow{1}{*}{Ours} 
 & \Ours  & \textbf{0.1407} & \textbf{01478} & \textbf{0.1580} & \textbf{0.1773} & \textbf{0.0197} & \textbf{0.0242} & \textbf{0.0359} & \textbf{0.0539} & \textbf{0.1768} & \textbf{0.2068} & \textbf{0.3124} & \textbf{0.4320} \\
\bottomrule
\end{tabular}
}
\vspace{-0.15in}
\end{table*}

For all three datasets, we arrange the interaction sequences in sequential order. We utilize a leave-one-out approach to split the data into training, validation, and testing sets. Detailed statistics of the datasets are provided in Table \ref{tab:1_data_statistics}.  The evaluation metrics are Normalized Discounted Cumulative Gain 
 (NDCG@K), Recall (Recall@K), which are evaluated on the full amount of data. The abbreviations N, and R are respectively used to denote NDCG, and Recall.

\subsubsection{Implementation Details}
We employ Qwen2vl-2b\footnote{\url{https://github.com/QwenLM/Qwen2-VL}} as the backbone model for both MIRM and DUEG (experiments with other MLLM backbones are presented in the appendix \ref{Impact_of_Different_MLLM_Backbone}). For each dataset, we create three types of data mixtures, each consisting of 10,000 data points, to fine-tune the MIRM.  Additionally, we employ SASRec as the ID-based recommendation model for contrastive learning, with an embedding dimension same as the MIRM.

For all methods involving LLMs, each experiment is trained for a maximum of 5 epochs with a batch size of 128. A learning rate warm-up strategy is employed, initializing the learning rate at 1/100 of its maximum value 1e-4, and dynamically adjusting it over training steps using a cosine scheduler.

\subsection{Performance Comparison (RQ1)}
In this section, we compare \Ours against traditional, content-based, and llm-based baselines, taking into metrics of both NDCG and Recall on PixelRec, MovieLens, and Amazon datasets, to showcase the effectiveness and robustness of \Ours.

\paragraph{Baselines.} %~{}

FPMC \cite{rendle2010factorizing}, GRU4Rec \cite{tan2016improved}, and SASRec \cite{kang2018self} are traditional sequential recommendation models based on Markov Chains, RNN, and attention mechanisms, respectively. DuoRec \cite{qiu2022contrastive} employs contrastive learning to extract discriminative information for sequential recommendation. SASRec-Content is a variant of SASRec that directly utilizes content feature representations as sequence inputs. It includes three versions: text-only, image-only, and a combination of text and image. CoLLM \cite{zhang2023collm} and HLLM \cite{chen2024hllm} are sequential recommendation models based on large language models (LLMs), both achieving state-of-the-art performance.

\paragraph{Main Results.}
We implemented the \Ours framework on three datasets. The comparison with baseline models is summarized in Table \ref{tab:2_main_results}. Key observations are as follows:

\noindent \textbf{(a) Superior Performance Across Datasets.} \Ours consistently outperforms all baseline models across three datasets. Our method achieves over a 7\% improvement in NDCG and Recall metrics on the MovieLens dataset, with similar enhancements on PixelRec and Amazon datasets compared to the strongest baseline. This demonstrates \Ours effectively integrates traditional sequential information with the expansive knowledge and reasoning capabilities of LLMs. By leveraging user interaction sequences, \Ours retains the strengths of collaborative filtering to capture user behavior patterns, while using MLLMs' advanced visual and language understanding to interpret complex user intents and contextual nuances. The synergy between sequential data and LLM-driven insights allows \Ours to balance explicit user preferences with implicit, semantically rich information, enhancing recommendation relevance and accuracy.

\noindent \textbf{(b) Enhanced Multimodal Integration.} By efficiently integrating textual and visual features through MLLM, \Ours achieves a substantial improvement in recommendation quality. Unlike SASRec-Content, which separately uses Vision Transformers (Vit) \cite{dosovitskiy2020image} for images and Bert \cite{devlin2018bert} for text, \Ours leverages the MLLM as MIRM. This integrated approach results in a remarkable 44\% improvement due to better item embedding extraction compared to SASRec-Content. The primary reason for this boost is the pretraining capabilities of MLLMs, which align and fuse visual and textual representations. This alignment allows for a deeper understanding of item features, streamlining the information processing pipeline and significantly enhancing the quality of the embeddings, thus improving recommendation performance.

\noindent \textbf{(c) Combining Semantics with Collaboration.} While methods like HLLM and CoLLM improve on ID-based and content-based approaches, they fall short compared to \Ours, particularly in NDCG metrics, due to their inability to capture collaborative knowledge essential for recommendations. LLMs excel in processing textual data but struggle with user-user and item-item interactions found in traditional systems. \Ours bridges this gap by combining LLMs' semantic understanding with the collaborative filtering strengths of traditional methods, enabling it to use both deep semantic insights and relational knowledge for more accurate and relevant recommendations.

\subsection{Impact of DUEG (RQ2)}

\begin{figure}[t] % 浮动环境
    % \vspace{-0.1in}
    \centering
    \includegraphics[width=0.5\textwidth, keepaspectratio]{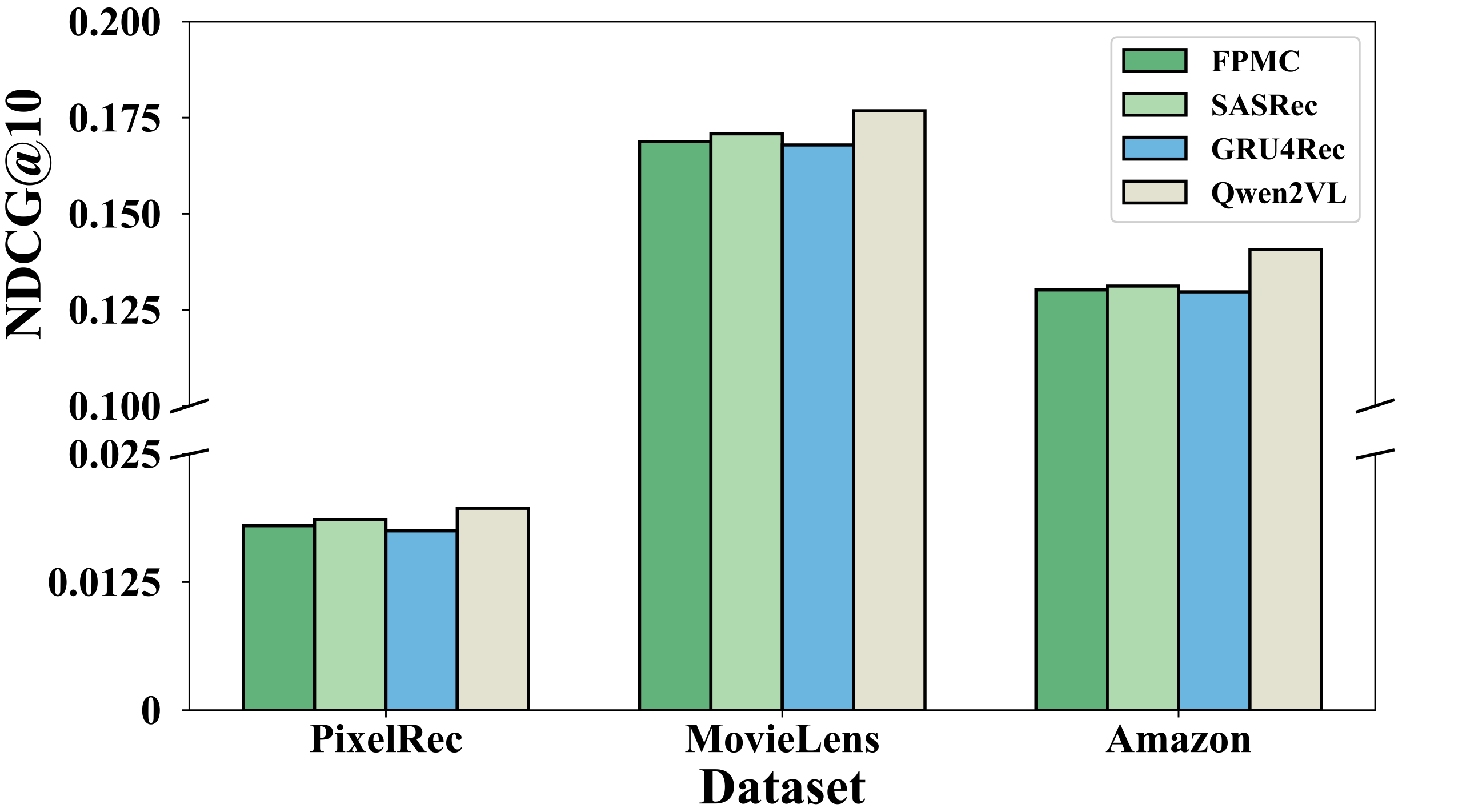} % 调整宽度        
    \vspace{-0.3in}
    \caption{\textbf{Performance comparison of different DUEGs.} Qwen2vl-2b is used as MIRM for all. The LLM backbone DUEG outperforms traditional DUEGs.  } % 添加标题
    \label{fig:2_GURM} % 添加标签用于引用
    \vspace{-0.2in}
\end{figure}

We conducted experiments to evaluate various representation methods for DUEGs, including FPMC, SASRec, GRU4Rec, and our proposed LLM Qwen2vl backbone as the DUEG.

As shown in Figure \ref{fig:2_GURM}, the results indicate that \Ours, using an LLM backbone as the DUEG, outperforms all other methods across the three datasets. This not only validates the effectiveness of our innovative user representation approach but also highlights the limitations of relying solely on textual and visual information (e.g., text and image metadata) or sequential information (e.g., behavioral ID tokens). Compared to the highest-performing baseline, SASRec, \Ours demonstrates an average improvement of 6.0\% on the PixelRec dataset and an even more substantial average improvement of 7.2\% on the Amazon dataset. The superior performance of \Ours can be attributed to the extensive pretraining of the LLM backbone, which imbues it with comprehensive world knowledge. Additionally, the alignment between pretraining, fine-tuning, and recommendation systems allows the training process to converge with only a few epochs (5 epochs), whereas other non-LLM baselines require prolonged training periods (e.g., 200 epochs for SASRec) to achieve convergence. These results show the superiority of the \Ours architecture in both performance and training efficiency.

% As shown in Figure \ref{fig:2_GURM}, \Ours, using an LLM backbone as the GURM, outperforms all other methods across the three datasets. This validates our innovative user representation approach and highlights the limitations of relying solely on textual, visual, or sequential information. Compared to the top baseline, SASRec, \Ours shows an average improvement of 6.0\% on the PixelRec dataset and 7.2\% on the Amazon dataset. The superior performance of \Ours stems from the extensive pretraining of the LLM backbone, which provides comprehensive world knowledge. Furthermore, the alignment between pretraining, fine-tuning, and recommendation systems allows the training process to converge with only a few epochs (5 epochs), while other non-LLM baselines need more extended training periods (e.g., 200 epochs for SASRec) to converge. These results demonstrate the superiority of the \Ours architecture in both performance and training efficiency.

\subsection{Impact of Input Data Modality (RQ3)}
To gain a thorough understanding of how various data modalities influence the performance of \Ours, particularly the contribution of multimodal fusion and the integration of complementary knowledge from various modalities, we conducted an in-depth analysis on PixelRec, showing in Table \ref{tab:3_modalities}.

\begin{table}[t]
% \vspace{-0.1in}
\centering
\caption{\textbf{Performance comparison with different modality inputs.} The table highlights the impact of using Image Only, Text Only, and Image + Text inputs for sequential recommendation tasks. The combined modality (Image + Text) consistently achieves the best performance across all evaluation metrics, demonstrating the advantage of multimodal integration.}
\label{tab:3_modalities}
{
\scalebox{0.7}{
\begin{tabular}{@{}lllllll@{}}
\toprule
 & N@10 & N@20 & N@50 & R@10 & R@20 &R@50 \\
\midrule
Image Only & \underline{0.0182} & 0.0217 &0.0292 &0.0329 & 0.0512 & 0.0858\\
Text Only &0.0181 & \underline{0.0228} & \underline{0.0296} & \underline{0.0335} & \underline{0.0514} & \underline{0.0860} \\
Image + Text & \textbf{0.0197} & \textbf{0.0242} & \textbf{0.0313} &\textbf{0.0359} & \textbf{0.0539} &\textbf{0.0895} \\
\bottomrule
\end{tabular}
}
}
% \vspace{-0.2in}
\end{table}

Our findings demonstrate that multimodal fusion enhances recommendation performance. The fusion of multiple modalities significantly enhances recommendation performance. A comparative analysis of \Ours on three input types reveals that the combined text with image input yields the best results. This can be attributed to the unique knowledge contributed by each modality, which cannot be captured by the others. MLLM effectively integrates this complementary information, demonstrating its potential as a robust foundation for multimodal recommendation tasks.

\subsection{Impact of Post-Alignment Models (RQ4)}
\begin{table}[t]
\centering
\caption{{\textbf{Performance comparison of different post-alignment models for contrastive learning.} Results show that stronger sequential models yield better performance, demonstrating the benefits of post-alignment.}}
\label{tab:6_post-alignment}
{\scalebox{0.8}{
    \begin{tabular}{clllll}
    \toprule
    Post-Alignment Model & N@10   & N@20    & R@10   & R@20     \\
    \midrule
    FPMC                 & 0.0194  & 0.0237 & 0.0347 & 0.0527  \\
    GRU4Rec              & 0.0195 & 0.0240  & \underline{0.0360} & 0.0531  \\
    SASRec               & \underline{0.0197} & \underline{0.0242} & 0.0359 & \underline{0.0539}  \\
    DuoRec               & \textbf{0.0200}   & \textbf{0.0253} & \textbf{0.0371} & \textbf{0.0569}  \\
    \bottomrule
    \end{tabular}
}
}
\vspace{-0.2in}
\end{table}

In the process of post-alignment contrastive learning, integrating ID information results in varying degrees of improvement across different traditional sequential recommendation models. To verify the performance impact brought by different traditional sequential recommendation models, we conducted the experiments shown in Table \ref{tab:6_post-alignment}.

The experimental results reveal a clear trend: stronger sequential recommendation models like DuoRec better support post-alignment contrastive learning, enhancing the integration of ID-based collaborative filtering signals into LLMs. This integration significantly boosts recommendation accuracy, coverage, and performance. DuoRec’s robust architecture captures richer user-item interaction patterns, enabling the LLM to leverage nuanced ID information for top NDCG and Recall scores. These findings highlight the importance of selecting powerful sequential models for contrastive learning, as they refine the process and ensure coherent ID integration, unlocking the full potential of collaborative filtering for diverse scenarios.

% without SA 1
% without IT 2
% without UB 3
% without contrastive learning  1.5
\begin{table}[t]
\centering
\caption{\textbf{Ablation study on the PixelRec dataset.} The table evaluates the impact of different fine-tuning data components (Image-Text, Structured Attributes, User Behavior) and the post-alignment module. Results demonstrate that using all fine-tuning components achieves optimal performance, while removing any single component degrades performance. The post-alignment contrastive learning module is shown to be critical for maintaining high recommendation accuracy.}
\label{tab:4_ablation_study}
{
\scalebox{0.68}{
\begin{tabular}{@{}llllllll@{}}
\toprule
& N@10 & N@20 &N@50 & R@10 & R@20 & R@50 \\
\midrule
\multicolumn{7}{c}{\textit{Full Model}} \\
\midrule
\textbf{\Ours} & \textbf{0.0197} & \textbf{0.0242} & \textbf{0.0313} & \textbf{0.0359} & \textbf{0.0539} & \textbf{0.0895}\\
\midrule
\multicolumn{7}{c}{\textit{Fine-Tuning Data}} \\
\midrule
\textit{w/o IT} & 0.0186 & 0.0227 & 0.0298 & 0.0339 & 0.0512 & 0.0841\\
\textit{w/o SA} & 0.0189 & 0.0237 & 0.0302 & 0.0349 & 0.0528 & 0.0859\\
\textit{w/o UB} & 0.0183 & 0.0220 & 0.0287 & 0.0324 & 0.0495 & 0.0828 \\ 
\textit{w/o ALL} & 0.0180 & 0.0219 & 0.0285 & 0.0313 & 0.0479 & 0.0808 \\
\midrule
\multicolumn{7}{c}{\textit{Post-Alignment}} \\
\midrule
\textit{w/o CL} & 0.0182 & 0.0225 & 0.0294 & 0.0325 & 0.0496 & 0.0819  \\
 
\bottomrule
\end{tabular}
}
}
\vspace{-0.2in}
\end{table}

\subsection{Ablation Study (RQ5)}

To analyze the contributions of components in the proposed \Ours method, particularly the fine-tuning of the MIRM on multimodal data and the post-alignment for user representation, we conducted an ablation study on the PixelRec dataset.

\paragraph{Effect of Fine-Tuning Data for MIRM.}  
The fine-tuning data for MIRM comprises three key components:
\begin{itemize}[left=0em, itemsep=-5pt, topsep=5pt]
    \item \textbf{Image-Text (IT):} Input: Item images. Output: Detailed textual descriptions.
    \item \textbf{Structured Attributes (SA):} Input: Item title and metadata. Output: Detailed descriptions.
    \item \textbf{User Behavior (UB):} Input: Historical item interactions (descriptions and images). Output: Predicted subsequent items.
\end{itemize}

To evaluate the contribution of each fine-tuning data component, we systematically removed one type of fine-tuning data at a time during stage 1. Additionally, removing all three types (\textit{w/o ALL}) effectively disables the fine-tuning stage. The results, presented in Table \ref{tab:4_ablation_study}, show that the model achieves its best performance when all three fine-tuning data types are used together. Conversely, removing any single type of data leads to a noticeable decline in performance, highlighting the importance of each component in enhancing the model’s overall effectiveness. Interestingly, the performance when structured attributes (\textit{w/o SA}) are excluded remains higher than when either image-text (\textit{w/o IT}) or user behavior (\textit{w/o UB}) data are omitted. This indicates that, even in the absence of structured attributes, the combination of image-text and user behavior data can effectively fine-tune MIRM, enabling it to learn and align meaningful representations.

\paragraph{Impact of Post-Alignment Contrastive Learning.}  
To examine the role of post-alignment in user representation, we conducted an ablation experiment by removing the post-alignment contrastive learning (\textit{w/o CL}) module. As shown in Table \ref{tab:4_ablation_study}, removing the contrastive learning module significantly decreases performance, particularly in the NDCG metric. This highlights the critical importance of incorporating ID-based information and collaborative signals. Post-alignment ensures effective feature-level communication and interaction between ID-based models and content-based MLLM, enabling the model to fully leverage multimodal information in a collaborative filtering context.

The ablation study highlights the importance of each fine-tuning data component and the post-alignment contrastive learning module. Fine-tuning on multimodal data ensures comprehensive item representations, while post-alignment bridges the gap between ID-based and content-based models, significantly enhancing the overall performance of the \Ours framework.

% As shown in Table \ref{tab:4_ablation_study}, the performance of the model deteriorates after removing the post-alignment contrastive learning (\textit{w/o CL}) module, especially in the NDCG metric. The results underscore the critical importance of incorporating ID information and collaborative signals. These elements ensure effective feature-level communication between ID-based models and content-based MLLM in a multimodal setting.

\section{Conclusion}
In this paper, we introduced \textbf{\Ours}, a novel framework for sequential recommendation that bridges the gap between collaborative filtering and multimodal content modeling using large language models (LLMs). While traditional LLM-based approaches excel in semantic understanding, they lack the ability to incorporate collaborative filtering signals, limiting their recommendation performance. To overcome this limitation, \textbf{\Ours} integrates multimodal data (textual and non-textual) with ID-based collaborative signals, leveraging an MLLM to generate unified item representations and a post-alignment mechanism to align user embeddings effectively.
By combining the strengths of multimodal content modeling and collaborative filtering, \textbf{\Ours} captures both user interests and contextual semantics, enabling precise and personalized recommendations. Extensive experimental results demonstrate that \textbf{\Ours} consistently outperforms traditional methods and state-of-the-art LLM-based baselines, validating its ability to fully exploit multimodal data and collaborative signals for sequential recommendation tasks.

% Limitation 可以在 conclusion 后面超过 8 页。
% \noindent\textbf{Limitation.}  
\paragraph{Limitation.}
While \Ours effectively integrates multimodal large language models (MLLMs) into sequential recommendation tasks, several limitations remain. First, the framework requires multi-task fine-tuning to optimize multimodal representations, which can be time-intensive and may hinder its deployment in real-time applications. Second, due to computational constraints, we are unable to train larger language models, and the quality of the generated multimodal item representations heavily depends on the underlying capabilities of the MLLMs. If the base models are suboptimal, the overall recommendation performance may degrade. In future work, we aim to develop an end-to-end training framework and incorporate more advanced MLLMs with larger parameter sizes to enhance the quality of generated representations, thereby improving the overall efficacy of \Ours.

\section*{Acknowledgement}
This research was supported by grants from the Provincial Natural Science Foundation of Anhui Province (No. 2408085QF193) and the Fundamental Research Funds for the Central Universities of China (No. WK2150110032, No. PA2024GDSK0112).

\bigskip
\bibliography{ref}

\cleardoublepage

\appendix

\section{Impact of Different MLLM Backbone}
\label{Impact_of_Different_MLLM_Backbone}
\begin{table}[h]
\centering
\caption{Comparison of Different MLLM Backbone.}
\label{tab:5_appendix_mllm_backbone}
{\scalebox{0.6}{
    \begin{tabular}{llllll}
    \toprule
    MLLM Backbone        & Training Type & N@10   & N@20    & R@10   & R@20     \\
    \midrule
    Qwen2-VL-2B          & Full-tuning   & \underline{0.0197} & 0.0242 & \underline{0.0359} & 0.0539  \\
    InternVL2.5-2B \footnote{\url{https://huggingface.co/OpenGVLab/InternVL2_5-2B}}       & Full-tuning   & 0.0191 & 0.0237 & 0.0349 & 0.0521  \\
    deepseek-vl-1.3b \footnote{\url{https://huggingface.co/deepseek-ai/deepseek-vl-1.3b-chat}}     & Full-tuning   & 0.0183 & 0.0225 & 0.0334 & 0.0499  \\
    \midrule
    Qwen2-VL-7B          & LoRA          & \textbf{0.0200}   & \textbf{0.0251} & \textbf{0.0369} & \textbf{0.0555}  \\
    Llama-3.2-11B-Vision \footnote{\url{https://huggingface.co/meta-llama/Llama-3.2-11B-Vision-Instruct}} & LoRA          & 0.0194 & \underline{0.0249} & 0.0357 & \underline{0.0542}  \\
    \bottomrule
    \end{tabular}
}}
\end{table}

To evaluate the impact of different MLLM backbones on the performance of \Ours, we conduct comparative experiments using MLLMs of various backbones and sizes. Due to computational constraints, we are unable to fine-tune models with 7B parameters or larger in full, so we employ LoRA training for models with 7B parameters and above.

As shown in Table 5, for models of the same size, performance variations across different backbones are observed, with Qwen2vl achieving the best results. This suggests that, beyond model size, the choice of backbone plays a crucial role in determining the quality of the recommendations. Interestingly, as the model size increases, there is a consistent improvement in recommendation performance, highlighting the advantages of scaling up model capacity. Even when fine-tuned with LoRA, the 7B Qwen2vl model consistently outperforms the 2B counterpart, indicating that the larger model not only benefits from increased parameters but also capitalizes on the specific architectural strengths of Qwen2vl. These findings suggest that, while model size is an important factor, selecting an appropriate backbone could be equally crucial in optimizing performance, particularly when computational resources are limited.

\section{Detailed Point-wise Recommendation Loss.}
\label{rec_loss}
To enhance the model's accuracy in predicting the next item, we employ a binary cross-entropy (BCE) loss function. In our training process, each positive sample is paired with a negative sample using a 1:1 negative sampling strategy. The target item embedding calculated from MIRM consists of a positive item (\(\text{pos}\)) and a negative item (\(\text{neg}\)). For each pair, we define label vector \(y = [1, 0]\) and generate predicted logits \(x = [x_{\text{pos}}, x_{\text{neg}}]\). The BCE loss is calculated as:

\vspace{-0.2in}
\begin{equation}
\mathcal{L}_{\text{bce}} = - \left( y \cdot \log(x) + (1 - y) \cdot \log(1 - x) \right)
\end{equation}

Minimizing this loss encourages the model to assign higher probabilities to positive samples and lower probabilities to negative samples, thereby improving its ability to distinguish relevant items for accurate next-item predictions.

\end{document}